\documentclass[12pt,preprint]{aastex}
\def\gtrsim{\mathrel{\hbox{\rlap{\hbox{\lower4pt\hbox{$\sim$}}}\hbox{$>$}}}}
\def\lesssim{\mathrel{\hbox{\rlap{\hbox{\lower4pt\hbox{$\sim$}}}\hbox{$<$}}}}
\newcommand{\msun}{$M_{\odot}$}
\newcommand{\ngc}{NGC~4254}
\newcommand{\VV}{VIRGOHI~21}
\newcommand{\hi}{H{\sc i}}

\newcommand{\swift}{{\sl Swift}}
\newcommand{\ergcms}{ergs~cm$^{-2}$~s$^{-1}$}
\newcommand{\fxfo}{$f_{\rm X}/f_{O}$}
\newcommand{\ergs}{ergs~s$^{-1}$}
\newcommand{\ergl}{ergs~s$^{-1}$}
\newcommand{\xrt}{\it XRT\rm}
\newcommand{\beq}{\begin{equation}}
\newcommand{\eeq}{\end{equation}}

\newcommand{\name}{SWIFT-XRT~J121723+1439.5}

\begin{document}

\title{Swift XRT Observations of the Possible Dark Galaxy \VV}
\author{Massimiliano~Bonamente\altaffilmark{1,2},
Douglas~A.~Swartz\altaffilmark{3},
Martin~C.~Weisskopf\altaffilmark{2}, and
Stephen~S.~Murray\altaffilmark{4}}
\altaffiltext{1}{Department of Physics, University of Alabama, Huntsville, AL, USA}
\altaffiltext{2}{NASA Marshall Space Flight Center, Huntsville, AL, USA}
\altaffiltext{3}{Universities Space Research Association,
    NASA Marshall Space Flight Center, VP62, Huntsville, AL, USA}
\altaffiltext{4}{Harvard-Smithsonian Center for Astrophysics, 60 Garden Street, 
MS-67, Cambridge, MA 02138}

\begin{abstract}
\swift\ \xrt\ observations of the \hi\ line source \VV\ were performed
on April 22 and April 26, 2008 for a total exposure time of 9.2~ks. 
This is the first pointed X-ray observation of \VV, a putative
dark galaxy in the Virgo cluster, and no photons were detected from this source. 
The non-detection of extended X-ray emission
within the angular extent of the \hi\ source
corresponds to a 99\% confidence upper limit of $2.1 \times 10^{-14}$~\ergcms\
in the 0.3--2.0~keV band.
The equivalent upper limit to the amount of diffuse hot gas
associated with \VV\ is in the range $4\times 10^{7} - 2 \times 10^8$  \msun\
for a hot gas temperature between 0.1 and 1 keV.
The non-detection also corresponds to a 99\%-confidence upper limit on the flux from
a point-like source 
of $8\times 10^{-15}$~\ergcms\ in the 0.3--2.0~keV
band.
We discuss the constraints on the nature of \VV\ imposed by these
observations and the theoretical implications of these results.
\end{abstract}


\section{Introduction}
\hi\ surveys of the Virgo Cluster \citep{davies2004,davies2006,
minchin2005,minchin2007,haynes2007}
have revealed a remarkable 
\hi\ cloud, \VV, with apparent evidence of rotation yet unaccompanied by any optical
emission. 
The \hi\ mass ($M_{HI} \sim 3\times 10^7$~\msun), 
inferred circular velocity ($V\sim 100$~km~s$^{-1}$) 
and physical size ($R\sim 8$~kpc) are all typical of luminous galaxies of 
dynamical masses $10^{10}-10^{11}$~\msun\ yet deep optical follow-up observations 
provide an upper limit of only 31.1 $I$ mag~arcsec$^{-2}$, corresponding to
a stellar content of 2$\times$10$^5$~$L_{\odot}$ \citep{minchin2007}. 
Thus, \VV\ is currently considered the 
prototypical dark galaxy candidate: a dynamically-massive 
dark-matter-dominated galaxy
 with a rotating gaseous disk but devoid of stars.

If \VV\ is  a massive dark galaxy, then it 
has important implications for the theory of structure formation.
The highly-successful cosmological hierarchical clustering model originated by 
\citet{white1978}  is based on the idea of accretion and merging of cold dark matter
(CDM) from small sizes ($\sim$10$^8$~\msun) up to the large galaxy clusters observed
at the present epoch. The CDM model predicts a large number of
$\gtrsim$10$^8$~\msun\ dark matter (DM) halos should remain unassimilated following
hierarchical growth and thus still be present today \citep[e.g.,][]{moore1999,
klypin1999,davies2006}.
To date, however, 
the number of observed low-mass galaxies falls far short of these 
predictions. This is true locally, where only a handful of satellites 
surround the massive Local Group spirals 
\citep[e.g.,][]{mateo1998},
and is manifest in
the faint end of the luminosity function that shows a slope more shallow than
$\Lambda$CDM predictions 
\citep{blanton2001,benson2003}.
This shortcoming of the $\Lambda$CDM model can be overcome either
by preventing small DM halos from forming in the first place
or by suppressing star formation so that they remain optically faint.
The latter possibility implies a population of
low-mass DM halos in 
which the baryon content remains optically-dark but still 
visible chiefly through \hi\ gas emission. 
The rarity of isolated extragalactic 
\hi\ clouds lacking optical counterparts 
\citep{doyle2005}
suggests that \VV, if truly massive, is a
rare case in favor of this scenario.

An alternative to the dark galaxy interpretation for \VV\ 
\citep{bekki2005, haynes2007, duc2008} is that
it is simply a part of an elongated \hi-dominated tidal
tail emanating from the nearby luminous galaxy \ngc\ (M99). The tidal tail was 
caused either by a fly-by encounter with another (massive) galaxy \citep{vollmer2005}
or by ram pressure stripping within the Virgo environment. 
Recent observations show the \hi\ gas in the tail extends from 
\ngc, located $\sim$120 kpc to the south of \VV, 
to about 130~kpc beyond \VV\ to the north \citep{haynes2007}.
The peculiar 
velocity structure in the vicinity of \VV\ is, in this case, ascribed to 
projection effects on the streaming motions of the debris tail 
instead of to rotation within a gravitationally bound disk.
In this scenario, the estimated \hi\ mass is close to the total mass of the object;
there is no massive halo accompanying \VV.

In 
\S~\ref{analysis} we describe the \swift\
observations leading to upper limits of the X-ray emission 
from \VV, and in \S~\ref{discussion} we provide our interpretation
of the \swift\ non-detection.
In this paper we assume 
a distance to the source of $D=16.5$ Mpc corresponding to the mean 
of the Virgo Cluster \citep{mei2007} and a scale of 4.9~kpc per arcminute.

\section{SWIFT observations of \VV\ and data analysis}
\label{analysis}

The \VV\ field was observed by \swift\ on 2008 April 22 and 26 with the
\xrt\ operated in photon counting mode.
The \xrt\ has a very low detector background, which makes it especially suitable for observations
of faint diffuse X-ray sources.
These data are the first pointed X-ray observations of \VV.

 The data were screened for bad pixels and other detector artifacts, and only photon
grades 0-12 were included in the analysis. The screening resulted in 9.2 ks
of clean data. In this analysis we considered photon energies in the 0.3-7 keV range,
where the calibration of the \xrt\ is better understood.
The data reduction was performed using \tt xselect \rm and the \tt FTOOLS\rm, 
and the spectral analysis with \tt XSPEC\rm.

\begin{figure}
\begin{center}
\includegraphics[angle=-90,width=6in]{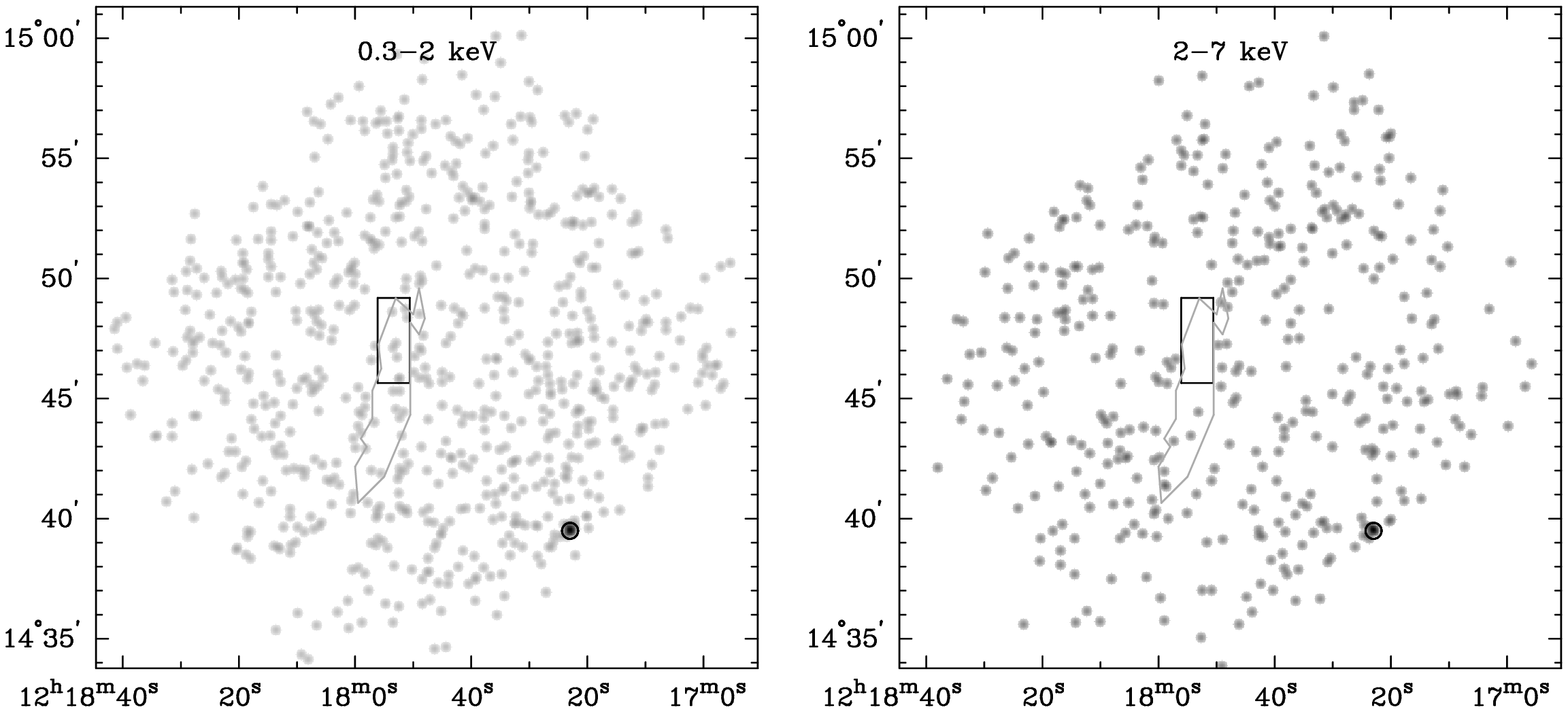}
\end{center}
\caption{\swift\ \xrt\ images of \VV \label{image}, smoothed with a Gaussian
kernel of $\sigma$=7". Grey contours are reproduced
from the \citet{minchin2007} \hi\ observations with the Westerbork telescope.
The black box represent the location of  the putative \hi\ rotating disc.
The circle centered at RA=12$^{h}$17$^{m}$23$^{s}$, $\delta$=14$^{\circ}$39'29" represents
the region from which we extracted the counts of the serendipitous source \name.
}
\end{figure}

\VV\ lies about 1 Mpc to the NW of the center of the Virgo Cluster
at R.A.$=$12$^{\rm h}$17$^{\rm m}$53.6$^{\rm s}$ 
decl.$=$+14\arcdeg45\arcmin25\arcsec\ (J2000.0). 
The angular size of \VV, 3.5\arcmin$\times$1.375\arcmin, lies well within
the 24\arcmin$\times$24\arcmin\ field of view of the \xrt\
\citep{cusumano2006,romano2005}.
X-ray images of the \VV\ field are shown in
Figure \ref{image}, in which the gray contour are from the \hi\ observations
of \citet{minchin2007}. The upper part of the \hi\ feature was interpreted 
by \citet{minchin2007} as a rotating disc, and the lower part as an \hi\ bridge connecting
\VV\ with NGC~4254, located to the south and outside the
\swift\ field of view. 

As with normal galaxies, \VV\ was expected to be a source of 
both diffuse and point-source X-ray emission.
Even in the absence of stars, 
gas falling into a galaxy's gravitational potential from the intergalactic medium
should be heated to roughly the (X-ray-emitting) virial temperature. 
In addition,
accretion onto a central supermassive black hole from the gas reservoir
in \VV, even at a fraction of the Bondi rate, should appear as a localized source of X-radiation though not necessarily optically bright.
No X-ray emission associated with \VV\ was detected.
An unrelated weak (41 counts) point-like source was detected near
R.A.=12$^{h}$17$^{m}$23$^{s}$, decl.=14$^{\circ}$39\arcmin29\arcsec\
(Figure \ref{image}).
As no object is known within 1\arcmin\ of this location,
we regard this as the discovery of the X-ray source \name.
In the following, we place upper limits to the presence of diffuse
and point source emission from \VV.

\subsection{Diffuse X-ray emission}
\label{diffuse}
In order to place upper limits to the diffuse soft X-ray emission 
from \VV, we accumulated X-ray photons from a 3.5'$\times$1.375'  
region located at the position of the putative \hi\ disc (black box 
in Figure \ref{image}). We detected a total of 8 counts in the 0.3-2 keV
band, and no counts in the 2-7 keV band. 
We estimate the number of background counts, first by extracting the data 
from the entire field of view, excluding a region of enhanced emission
to the south (Figure \ref{image}) and regions close to the detector
boundaries. The region included encompassed 337.3 square arcminutes. 
We found $623$ counts in the 0.3-2 keV band and $319$ counts in the 2-7 keV 
band. Ignoring the possibility that some of these events might be vignetted by 
the telescope one therefore expects in the region of interest at least 
$8.9\pm0.4$ counts in the soft band and $4.6\pm0.3$ in the hard band. We 
note that it is mildly interesting and somewhat improbable that no counts 
were measured in the hard band from the region containing \VV.

We use the soft X-ray band data in order to set upper limits as to source 
detection, since this is the band in which
the X-ray emission from diffuse gas (and point sources) is expected.
Since the measured background predicts at least a mean of 8.9$\pm$0.4 counts 
and we detected 8 counts it is therefore evident that the \VV\ region is not 
an X-ray emitter.
To determine upper limits  we increased the expected number of counts in the 
soft band by 10\% to 9.8 counts to account for the possibility that some of 
the measured background events arise from faint, unresolved X-ray sources.
We determine 99\%-confidence upper limits assuming 18 counts as detection.

\subsubsection{Uniform distribution}
\label{sec:uniform}
\begin{figure}
\includegraphics[angle=-90,width=5.0in]{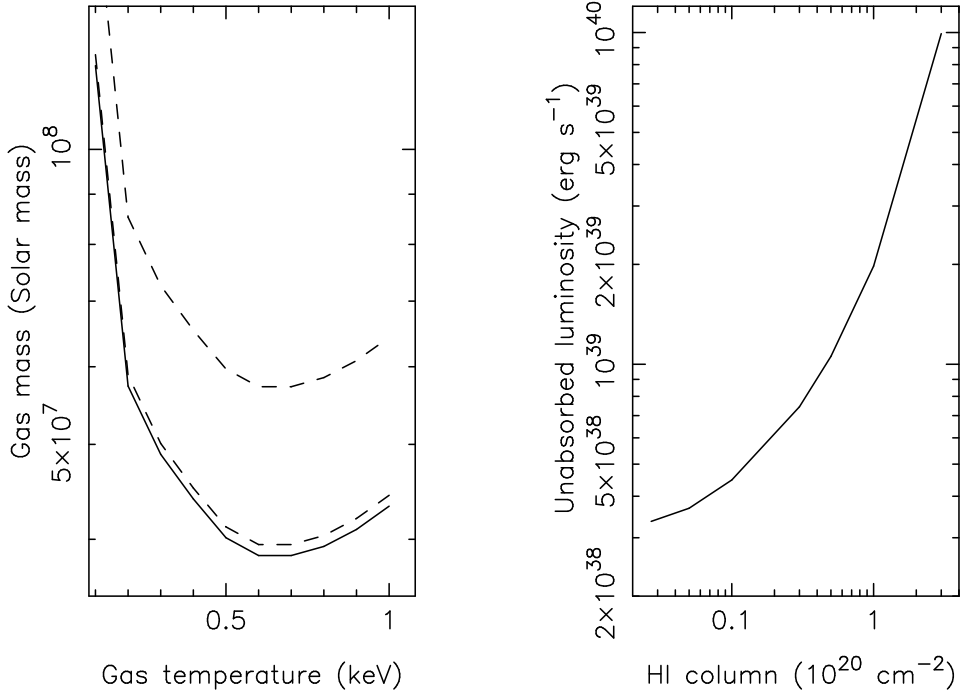}
\caption{(a): Upper limits to the mass of diffuse gas in \VV, estimated assuming a 
detection threshold of 17 counts in these \xrt\ observations. The \swift\ \xrt\ bandpass
is not sensitive to emission from diffuse gas at temperature $kT \leq 0.1$ keV.
Solid line : uniform
density model, using a detection threshold of 8.2 counts in a 3.5'$\times$1.375' box.
Dashed lines : $\beta$ model with $r_c$=0.5', using a detection threshold
of 10.7 counts in a circle of radius 1.75' (bottom line: $\beta$=0.67; top line: $\beta$=0.33).
Given that the gas mass is proportional 
to the square root of the number of detected counts, the estimates are not very sensitive
to changes in the detection threshold.
(b) Upper limits to the luminosity of point sources in the \VV\ field, estimated
assuming a detection threshold of 4 source counts. The luminosity is proportional to the
number of detected counts.
\label{limits}}
\end{figure}

We assume that the \hi\ disk is seen edge-on, and thus the emitting volume
is $V=\pi R^2 L$, where $R=1.75$' and $L=1.375$'. 
 We use an optically-thin plasma 
of uniform density and of Solar abundance (\tt apec \rm in \tt XSPEC\rm), a Galactic
\hi\ column density of $N_H=2.7\times 10^{20}$ cm$^{-2}$
\citep{dickey1990,kalberla2005} with \citet{morrison1983} cross-sections (\tt wabs \rm
in \tt XSPEC\rm), and calculate the emission measure
required to achieve 8.1 source counts as a function of plasma temperature using \tt XSPEC\rm.
The emission measure calculated by \tt XSPEC \rm is given by the model normalization, $K$, as
\begin{equation}
K= \frac{10^{-14}}{4 \pi D_A^2} \int n_e n_H dV,
\label{norm}
\end{equation}
in which $n_e$ and $n_H$ are, respectively, the electron and hydrogen
number density and $D_A$ the angular size distance (for $z\ll 1$).
From this, we calculate the gas density, and thus the gas mass.
The results are shown in Figure \ref{limits}, showing that these 9.2 ks \swift\
\xrt\ observations set upper limits of the order of $10^8 M_{\odot}$.
These upper limits correspond to a flux of $2.1 \times 10^{-14}$ \ergcms\ in the 0.3-2 keV band.

\subsubsection{$\beta$ model distribution}
\label{sec:beta}
A putative hot halo may be maintained in hydrostatic equilibrium by the
gravitational potential of the galaxy. In this case, 
the gas density will decrease with radius. We therefore 
consider a more realistic model of the gas density
given by the $\beta$ model,
\begin{equation}
n_{e}=n_{e0} \left( 1 + \frac{r^2}{r_c^2} \right)^{-\frac{3}{2} \beta}
\label{beta}
\end{equation}
as often observed in galaxy clusters \citep{cavaliere1978}. The parameters
$r_c$ and $\beta$ describe, respectively, the radius within which
the density is approximately constant, and the decrease
of the density at large radii. For this interpretation, we use a spherically
symmetrical model in which the galaxy has a radius of $R=1.75$'.
In this region, the background in the 0.3-2 keV band has a mean of
16.3 counts, and we thus place the 99\% confidence upper
limit at 27 total counts (or 10.7 source counts), following the method described in
\S~\ref{sec:uniform}. Using the $\beta$ model (Equation \ref{beta})
into Equation \ref{norm}, we obtain upper limits to the central
gas density:
\begin{equation}
n_{e0}=\left(\frac{K 10^{14} 4 \pi D^2}{4 \pi r_c^3 I_1(R/r_c)}\right)
^{\frac{1}{2}}
\label{neo}
\end{equation}
in which the integration of the density profile has led to the integral
$I_1(R/r_c) = \int_{0}^{R/r_c} x^2(1+x^2)^{-3 \beta} dx$.
We use two reference values for the index $\beta$: 
a value of $\beta$=2/3, as usually measured in galaxy
clusters \citep[e.g.,][]{laroque2006}, 
and a shallower slope  of $\beta$=1/3, 
which may be typical of lower-mass halos (e.g., galaxy groups
and galaxies).
We find analytical solutions
for the integral, $I_1(n)=n/(2+2n^2)+ \arctan(n)/2$ for $\beta$=2/3,
and $I_1(n)=n-\arctan(n)$ for $\beta$=1/3. 

We then estimate the upper limits to the gas mass by integration of the
density profile,
\begin{eqnarray}
M_{gas} &=& \mu_e m_H \int_0^{R} n_{e0} \left( 1 + \frac{r^2}{r_c^2} \right)^{-\frac{3}{2} \beta}
4 \pi r^2 dr \nonumber \\ &=& 4 \pi \mu_e m_H n_{e0} r_c^3 I_2(R/r_c),
\label{mgas}
\end{eqnarray}
in which the $\beta$ model profile has led to the integral
$I_2(R/r_c) = \int_0^{R/r_c} x^2(1+x^2)^{-3\beta/2} dx$.
This integral is $I_2(n) = n - \arctan(n)$ for $\beta=2/3$, and  $I_2(n)=\ln(x+\sqrt{1+x^2})$
for $\beta=1/3$.

We estimate upper limits to the gas mass
as function
of plasma temperature using Equations \ref{neo}  and \ref{mgas},
using a fiducial value of the core radius of 0.5' (Figure \ref{limits}).
We determine that the upper limits to the gas mass depend weakly on the choice of
the core radius, given that varying the fiducial value by $\pm$100\% results in 
variation of less than 10\%.  
This model results in upper limits to the gas mass that are in the range
of $M_{gas} \simeq 4\times 10^7 - 2 \times 10^8$ $M_{\odot}$.
Upper limits to the gas mass are therefore similar to those obtained from
the simple uniform density model of Section \ref{sec:uniform}. 

\subsection{Point source X-ray emission}
\label{ps}
We set upper limits to the luminosity of an unobserved point source anywhere 
within the 3.5' x 1.375' region of interest. The upper limit takes into account 
that there are 68 statistically-independent SWIFT spatial resolution elements 
within this region, given that the half-power diameter of the \xrt\ point spread function
at 1.5 keV is 18". The measured background, per 18"-diameter spatial 
resolution element is 0.131, or 0.144 using a +10\% allowance for 
vignetting. Thus, considering the Poisson distribution with a 
mean of 0.144 and a total of 4 counts, there is a probability of only $1.6\times 10^{-5}$ of 
measuring 4 or more counts in just one of these resolution elements. Of 
course we would have been happy to measure 4 or more counts in any one of the 
68 elements so the joint probability is correspondingly larger. Thus, we may 
state, with approximately 99.9\%-confidence that we have not detected a 
a source of 4 counts or brighter from this region.

We consider a power-law emission model with photon index $\Gamma=2.0$
and determine the upper limits to the source luminosity as function
of the absorbing \hi\ column density using \tt XSPEC\rm. 
The upper limits are shown in Figure \ref{limits},
and are in the $L=3\times 10^{38}-10^{40}$ \ergl\ range. This upper limit 
corresponds to a flux of 8$\times 10^{-15}$ \ergcms\ in the 0.3-2 keV
band.
 
\subsection{UV and IR emission}

For completeness, we also analyzed extant non-X-ray images of \VV.
 
\swift\ UVOT images using the UVM2 filter 
(effective wavelength 2231 \AA) were obtained 
concurrently with the XRT data. No sources were detected within 
the angular extent of \VV. The {\tt FTOOLS} utility {\tt uvotsource}
was used to estimate a limiting magnitude for the UV emission from
the 3.5\arcmin$\times$1.375\arcmin\ region enclosing \VV.
This limit does not improve our estimates of the upper limit to 
the mass of diffuse hot gas nor to the flux from a point source
as estimated above from the X-ray data.

We also analyzed archival {\sl Spitzer} data (Program ID 30725)
which included all four IRAC and three MIPS spectral bands.
Again, no extended emission was detected within the \hi\ extent of
\VV. We estimate an upper limit to the mass of dust associated with
\VV\ to be 10$^4$--10$^6$~\msun\ depending on the dust temperature
assumed. 

\section{Discussion and conclusions}
\label{discussion}

A search for X-ray emission from the \hi\ line source \VV\ using the
\swift\ \xrt\ obtained only upper limits.
The non-detection indicates that any diffuse X-ray emitting gas at $kT=0.1-1$ keV
must be less massive than $4\times 10^{7} - 2 \times 10^8$  \msun, 
depending on the gas temperature and its spatial distribution. Given that the \swift\
\xrt\ detector has no effective area approximately below 0.1 keV, it is possible that
significant amounts of gas at sub-virial temperature have gone undetected by this
X-ray observation.
Similarly, any X-ray point source associated with \VV\ will have a 
luminosity lower than $5 \times 10^{38}-10^{40}$ \ergs, depending on the
absorbing column density of gas along the line of sight.

How do these limits compare with expectations? There are well-established
correlations between optical and X-ray light from galaxies: Typical
\fxfo\ ratios for normal galaxies range from $10^{-4}$ to $10^{-2}$ 
and for galaxies hosting AGNs the range extends up to unity 
\citep[e.g.,][]{kim2006}. 
For the dark galaxy \VV, these ratios predict an X-ray
flux of $10^{-17}$ to $10^{-15}$ for a normal galaxy up to $10^{-13}$
if an AGN is present. 
Of course, these \fxfo\ correlations reflect the causal link between
stellar processes and X-ray light. They are not expected to hold in the
absence of stars.

Instead, our
expectation that \VV\ should be a source of diffuse X-ray emission
stems from the fundamental picture \citep{reesostriker1977} that gas 
falling into a DM halo is first heated to the virial temperature. 
The accumulation of hot gas over a Hubble time could be substantial for
a massive DM halo like \VV\ \citep{whitefrenk1991}.
We would expect a gas virial temperature of 
approximately 0.4~keV based on the observed kinematics of \VV\ 
\citep[from][]{minchin2007} and thus our best guess to the 
amount of hot gas associated with \VV\ is $\leq 5\times 10^7$~\msun.
%
Scaling from simulations of gas infall onto stable disk galaxies
\citep{benson2000, toft2002},
the X-ray luminosity of this gas should be up to 
5$\times$10$^{38}$~\ergl\ for rotation speeds similar to that 
of \VV. 

The above estimates are based on the assumption of an isolated `quiescent' massive
galaxy accreting slowly (i.e., with a long cooling time) 
from the surrounding intergalactic medium. 
In the particular case of \VV, in contrast,
a considerable gas reservoir exists within the tidal tail emanating
from \ngc\ that should lead to a 
relatively high gas accretion rate and hence high X-ray luminosity.
Furthermore, turbulence and shocks following the interaction between
\VV\ and \ngc\ would very likely heat some portion of the disk \hi\ gas component 
of \VV\ to X-ray emitting temperatures.
We therefore consider the above estimates to be minimal expectations.
Of course, any prediction of the amount of hot gas or its luminosity 
will scale with the dynamical mass of 
the accreting object. 

The observed upper limits reported here, therefore,
show that the extreme object \VV\ is {\sl not} an extreme X-ray object 
(in the sense of a high \fxfo) nor has it accreted very much gas 
from its surroundings during its lifetime.

\clearpage

\end{document}